\documentclass[preprint,showpacs,aps]{revtex4}
\begin{document}
\draft
\title{A variational Jastrow coupled-cluster theory of quantum many-body systems}
\author{Y. Xian}
\address{School of Physics and Astronomy, The University of
Manchester, Manchester M13 9PL, UK} \date{\today }

\begin{abstract}
We study many-body correlations in the ground states of a general quantum system 
of bosons or fermions by including an additional Jastrow function in our recently
proposed variational coupled-cluster method. Our approach combines the advantages
of state-dependent correlations in the coupled-cluster theory and of strong, 
short-ranged correlations of the Jastrow function. We apply a generalized linked-cluster
expansion for the Jastrow wavefunction and provide detailed analysis
for practical evaluation of Hamiltonian expectation value as an energy functional
of the Jastrow function and the bare density-distribution functions introduced and 
calculated in our earlier publications; a simple, first-order energy functional
is derived and detailed formulas for higher-order contributions are provided.
Our energy functional does not suffer the divergence as in most coupled-cluster
calculations when applying to Hamiltonians with hardcore potentials.
We also discuss relations between our energy functional and the energy functionals
from other theories.
\end{abstract}
\maketitle

\section{introduction}

Most microscopic quantum many-body theories developed over the last
five decades can perhaps be broadly divided into two categories,
one in real space and the other in momentum (configurational) space.
While a real-space many-body theory usually focuses on the interaction
potential part of Hamiltonian and evaluates the Hamiltonian expectation
value in a first quantization form, a momentum-space theory often 
starts from the kinetic part of Hamiltonian, is a basis of or 
closely related to many-body perturbation theories, and mostly
deals with Hamiltonian in a second quantization form \cite{thou,ripk}.
A typical real-space approach to the ground state of a quantum many-body system 
is provided by Jastrow wavefunction which is constructed by a 
state-independent two-body correlation function \cite{jast}. Systematic
techniques based on Jastrow wavefunction including extension to
inhomogeneous boson or fermion systems is now generally referred to as the method
of correlated basis functions (CBF) \cite{feen, krot,fant}. The CBF method has
proved to be efficient in dealing with strong, short-ranged 
correlations typified by those in quantum helium liquids \cite{feen}. On the other hand,
momentum-space many-body theories are often easier to apply
and, due to inclusion of the state-dependent correlations, is capable of
producing accurate results for a wide range of quantum systems, such 
as boson gas \cite{bogo}, quantum antiferromagnets with N\'eel 
order \cite{ande}, finite nuclei \cite{day}, and electron systems such as 
electron gas \cite{bish}, atoms and molecules \cite{pald}. A typical momentum 
theory is the coupled-cluster method (CCM) in which wavefunction are explicitly 
constructed by state-dependent operators \cite{coes,cize}.
State-of-the-art calculations of the CCM with high accuracy have often been
carried out in quantum chemistry \cite{bart}, and recently in quantum 
spin lattices with N\'eel order \cite{bish2}. Systematic 
resummations of diagrams in perturbation theory
for boson systems have revealed interesting relations between the 
two approaches, for example, the hyper-netted chain approximation in
the CBF method in fact contains a consistent resummation of both infinite ring and 
infinite ladder diagrams of momentum-space approach \cite{ripk2,bish3}.
It appears that real-space and momentum-space approaches complement each other and
unification of these two approaches may provide a quantitative description
applicable to wider range of quantum many-body systems, including in particular
the strongly-correlated fermion systems \cite{krot2}.

We recently extended the CCM to a variational formalism in which bra and 
ket states are now hermitian to one another \cite{yx1,yx2,yx3}, contrast to the
traditional CCM where they are not \cite{arpo}. We introduced the 
hermitian-conjugate pair of important bare density distribution functions for 
practical and systematical calculations; the traditional CCM was shown 
to correspond to a simple linear approximation in one set of
distribution functions of our variational coupled-cluster method (VCCM). The 
well-known momentum approaches such as Bogoliubov theory of boson gas, 
Anderson's spin-wave theory (SWT), and BCS theory of superconductivity \cite{BCS}, 
can all be explicitly shown as special low-order approximations 
in both the ground- and excited-state wavefunctions of the VCCM.
We have demonstrated by a detailed application to quantum
antiferromagnets with N\'eel order. Approximations beyond
SWT improved results for the ground-state properties \cite{yx2} and new
excitation states have also been obtained \cite{yx3}. Furthermore,
our calculations for the bare density distribution functions can be carried 
out by diagrammatical techniques similar to those employed by the CBF 
methods \cite{yx2}. Hence, a bridge between coupled-cluster theory and Jastrow theory
is built. We therefore believe it is a natural next step to combine the two methods 
for a unified description. This is our main purpose in this article. Krotscheck, 
K\"ummel and Zabolitzky (KKZ) made the first attempt in 1980 for fermion
systems \cite{krot2}. They employed the traditional CCM with only the ket
state specified and the Jastrow function is fixed before hand; Hamiltonian
eigenequation was used to obtain the ground-state energy and equations 
for the ket-state coefficients. Here we employ explicit ket and bra states
of the VCCM and calculate distribution functions in terms
of these ket- and bra-state coefficients; the energy functional is derived
in terms of the distribution functions and the Jastrow function.

We focus here on formal development of our approach and organize 
this article as follows. In Sec.~II we introduce our correlated wavefunctions 
and their distribution functions. The generalized linked-cluster
expansion technique is employed for calculations of generating functional in
terms of the bare distribution functions. In Sec.~III we evaluate Hamiltonian
expectation value using Jackson-Feenberg transformation; a first-order energy
functional is derived and formulas for higher-order contributions are 
provided. We conclude in Sec.~VI with a summary and discussion on 
the relations between our approach and other many-body theories.

\section{Distribution functions in variational Jastrow coupled-cluster theory}

As proposed earlier \cite{yx2}, we consider a general variational wavefunction 
by adding a Jastrow correlation operator on top of the Coester state $|\Psi^c\rangle$
\begin{equation}
|\Psi\rangle =e^{S^0/2}|\Psi^c\rangle,
\end{equation}
and we use the following notation
\begin{equation}
\langle\tilde\Psi| =\langle\tilde\Psi^c| e^{S^0/2},
\end{equation}
for the bra state. In Eqs.~(1) and (2), the hermitian operator $S^0$ is the 
Jastrow correlation operator, and is given in terms of field operators as \cite{ripk},
\begin{equation}
S^0=\frac12\int dx_1dx_2\psi^\dagger(x_1)\psi^\dagger(x_2) u(x_1,x_2)
\psi(x_2)\psi(x_1),
\end{equation}
where field operators $\psi^\dagger(x)$ and $\psi(x)$ obey the usual boson or fermion
commutation relations, $u(x_1,x_2)$ is a local, symmetric function $u(x_1,x_2)=u(x_2,x_1)$,
and $x$ are particle coordinates including spin degrees of freedom.
We also require that $u(x_1,x_2)$ is bound and short-ranged in real space.
We do not include in $S^0$
any single-body operator as it can be easily absorbed in the Coester states to be
defined next. The boson or fermion symmetry is contained in 
the Coester states by definition. More specifically, the Coester ket state is 
given by, using the convenient notation invented by Arponen and 
Bishop \cite{arpo2},
\begin{equation}
|\Psi^c\rangle=e^S|\Phi\rangle,
\end{equation}
where $S$ is constructed by the so-called configurational creation operators
$C^\dagger_I$ which are defined with respect to the model state $|\Phi\rangle$ with
the nominal index $I$ labeling multi-particle excitation states from the model 
state $|\Phi\rangle$,
\begin{equation}
S=\sum_I F_I C^\dagger_I,
\end{equation}
with $F_I$ often referred to as the correlation coefficients.
The Coester bra-state $\langle\tilde \Psi^c|$ is given by the hermitian
conjugate of the ket state,
\begin{equation}
\langle\tilde\Psi^c|=\langle\Phi|e^{\tilde S},\quad \tilde S=\sum_I\tilde F_I C_I
\end{equation}
in our VCCM, where $C_I$ are the corresponding configurational
destruction operators and $\tilde F_I$ are the independent, hermitian conjugates of
$F_I$. In the traditional CCM, however, the bra state is parametrized 
differently from the ket state and is written as \cite{arpo},
\begin{equation}
\langle\tilde \Psi^c| =\langle\Phi|\tilde S' e^{-S},
\quad \tilde S' = 1+\sum_I\tilde F'_I C_I.
\end{equation}
where $S$ is defined as in the ket state of Eq.~(5), and $\tilde F'_I$ is the
bra state coefficients which, in general are not hermitian conjugate of $F_I$. 
As discussed in Appendix A, the CCM states violate the condition for application of
the generalized linked-cluster expansion due to the {\it linear} construction of 
the bra state. Furthermore, the evaluation of the kinetic energy
discussed in Appendix B will contain a three-body terms because 
the nonhermitian relation between the ket and bra states in the CCM.
We therefore will not discuss the CCM further
and focus only on the VCCM basis of Eqs.~(4-6). Clearly, the natural 
variational parameters are $(F,\tilde F,u)$ in the VCCM basis, where we have used the 
notations $F=\{F_I\}$, $\tilde F=\{\tilde F_I\}$ and $u=u(x_1,x_2)$. In principle, 
if these Coester states are exact (namely, all configurations are included in the 
summations over all $I$-indices), parameter $u$ is redundant. However, as we always 
need to make a finite truncation approximation in summations over
$I$-indices in any practical application and it is well-known that
the Coester states in a finite truncation approximation are not
efficient in dealing with the strong, short-ranged correlations, the
two-body Jastrow function $u(x_1,x_2)$ is a useful, important
variational parameter in a real application. We hence always 
assume the summations in Eqs.~(5) and (6) are within the subset of
a truncation approximation. One of such truncations is the so-called SUB$m$
approximation in which we retain up to $m$-body creation operators only.

Our basic strategy for calculations is to evaluate the generating 
functional $W$ of Eqs.~(1) and (2),
\begin{equation}
W=\ln \langle\tilde\Psi|\Psi\rangle = W^c + W^u,
\end{equation}
where $W^c$ is the generating function for the pure Coester states
without the Jastrow operator,
\begin{equation}
W^c = \ln \langle\tilde\Psi^c|\Psi^c\rangle,
\end{equation}
and $W^u$ is the remainder containing the $u$ function
\begin{equation}
W^u = \ln \langle e^{S^0} \rangle^c,\quad
\langle e^{S^0} \rangle^c = \frac{\langle\tilde\Psi^c|e^{S^0}|\Psi^c\rangle}
 {\langle\tilde\Psi^c|\Psi^c\rangle}.
\end{equation}
The general strategy for calculating $W^c=W^c(F,\tilde F)$ was discussed 
in our earlier papers and detailed calculations were demonstrated in the 
spin-lattice application \cite{yx1,yx2}. Briefly,
we first introduce the hermitian-conjugate pair of the bare density distribution 
functions
\begin{equation}
\tilde g^c_I =\langle C^\dagger_I\rangle^c =\frac{\partial W^c}{\partial F_I},\quad
g^c_I =\langle C_I\rangle^c =\frac{\partial W^c}{\partial\tilde F_I},
\end{equation}
where the expectation values $\langle\cdots \rangle^c$ are calculated using 
the Coester states of Eqs.~(4-6) as was given in Eq.~(10); any physical quantity
is expressed in terms of these distribution functions; these distribution 
functions are then calculated either by algebraic technique \cite{yx1} or by 
diagrammatic technique \cite{yx2}. In the algebraic approach
we derive the following general and self-consistency equations for the calculations,
\begin{equation}
\tilde g^c_I =G_I(\tilde F,g^c),\quad g^c_I = G_I(F,\tilde g^c)
\end{equation}
where $G_I(\tilde F,g^c)$ with notation $g^c=\{g^c_I\}$ etc. as before
is a function containing up to linear terms in $g^c$ and finite-order
terms in $\tilde F$, and similarly for $G_I(F,\tilde g^c)$. In the
diagrammatic approach, many techniques employed by the CBF method
are also applicable here. For example, the coefficients $\tilde F_I$ can be
replaced in every diagram of $W^c$ expansion by the bare distribution 
functions $\tilde g_I$ after resummations of infinite ring diagrams. In general,
the energy expectation value can be expressed as a functional polynomial
in $F$ and $\tilde g^c$, $E^c=E^c(F,\tilde g^c)$; and the variational 
equations can then be derived by the functional derivatives \cite{yx4}. 

In terms of the distribution functions as discussed above, we now consider 
evaluation of $\langle e^{S^0}\rangle^c$ in Eq.~(10). After cluster expansion
for the Jastrow part of the wavefunctions, we apply the generalized linked-cluster
theorem discussed in Appendix A in a similar fashion as the traditional
Jastrow theory \cite{ripk,clar}, and we obtain
\begin{equation}
W^u=\ln\langle e^{S^0}\rangle^c = \sum_{n=2}\frac1{n!}\int dx_1\cdots dx_n [Z_n \rho^c_n]_L
\end{equation}
where $Z_n(Y)$ are the $n$-body Yvon-Mayer functions of the bound function $Y_{12}$,
\begin{equation}
Y_{12}=e^{u(x_1,x_2)}-1;
\end{equation}
$\rho^c_n$ is the $n$-body density distribution functions of the Coester states;
and the notations $[\cdots]_L$ denote that only the linked components of the products are
included. In deriving Eq.~(13) in Appendix~A, we have assumed that the 
Coester states satisfy the cluster-decomposition properties (i.e., higher-order $\rho^c_n$
are given by sum of products of lower-order ones plus a nondecomposable core).
While this is true in a SUB2 approximation in our spin lattice application
and the SUB2 approximation for the Bose gas and BCS superconductors \cite{yx2,yx5},
we can not provide a general proof without the details of truncation approximations employed
in the Coester states in a real application. Hence by using Eq.~(13), we have made a further
restriction in the truncation approximations of the Coester states. In the following 
discussion this further approximation is understood and we will examine this 
property in real applications beyond the SUB2 approximation. The following
analysis for calculating $\rho^c_n$ in terms of $F_I$ and $\tilde g^c_I$ clearly
show such examination posing no major difficulty.

As all the Yvon-Mayer diagrams in $Z_2$ and $Z_3$ are linked, we simply have 
\begin{equation}
[Z_2\rho^c_2]_L=Y_{12}\rho^c_2,\quad 
\rho^c_2(x_1,x_2) =\langle\psi^\dagger(x_1)\psi^\dagger(x_2)\psi(x_2)\psi(x_1)\rangle^c
\end{equation}
for 2-body cluster contribution and
\begin{equation}
[Z_3\rho^c_3]_L=(Y_{12}Y_{23}+Y_{23}Y_{31}+Y_{31}Y_{12}+Y_{12}Y_{23}Y_{31})\rho^c_3
\end{equation}
for the 3-body cluster contributions with the 3-body distribution function
\begin{equation}
\rho^c_3 =\langle\psi^\dagger(x_1)\psi^\dagger(x_2)\psi^\dagger(x_3)
\psi(x_3)\psi(x_2)\psi(x_1)\rangle^c
\end{equation}
in the Coester states. From the 4-body function $Z_4(Y)$ and onward, 
however, there are unlinked Yvon-Mayer diagrams which are to be 
included in the product $[Z_n\rho^c_n]_L$ only after multiplying with the terms 
of $\rho^c_n$ to form the linked components, and details of which will depend
on applications with a truncation approximation 
in the Coester states. There is clearly a 
trade-off between the order of linked-cluster expansion and the order of 
truncation approximation of the Coester states. We hope to get experience 
in real applications in future. The additional state-dependent correlations
of the Coester states are contained firstly in $W^c$ of Eq.~(9) and secondly
in the density distribution functions $\rho^c_n$.
Importantly, these density distribution functions can be calculated 
in terms of $F$ and $\tilde g^c$, and by using the linearity theorem of the 
VCCM \cite{yx3}, we can show that all $\rho^c_n$ functions contain
only up to linear terms in $\tilde g^c$ and finite-order terms in $F$.
As a demonstration, we consider the two-body function $\rho^c_2$,
\begin{equation}
\rho^c_2(x_1,x_2) =\langle A_2\rangle^c=\frac1{\langle\tilde\Psi^c|\Psi^c\rangle}
  \langle\tilde\Psi^c|A_2e^S|\Phi\rangle=
  \frac1{\langle\tilde\Psi^c|\Psi^c\rangle}
  \langle\tilde\Psi^c|e^S\bar A_2|\Phi\rangle,
\end{equation}
where $A_2= \psi^\dagger(x_1)\psi^\dagger(x_2)\psi(x_2)\psi(x_1)$
and $\bar A_2$ is
\begin{equation}
\bar A_2 =e^{-S}A_2e^S =e^{-S}\psi^\dagger(x_1)\psi^\dagger(x_2)\psi(x_2)\psi(x_1)e^S
 = \bar\psi^\dagger(x_1) \bar\psi^\dagger(x_2) \bar\psi(x_2) \bar\psi(x_1),
\end{equation}
with $\bar \psi^\dagger(x) =e^{-S}\psi^\dagger(x)e^S$ etc.
Using the nested commutation series, we obtain
\begin{equation}
\bar \psi^\dagger(x) =\psi^\dagger(x)+[\psi^\dagger(x),S]
  +\frac1{2!}[[\psi^\dagger(x),S],S]+\cdots
  =\psi^\dagger(x)+[\psi^\dagger(x),S],
\end{equation}
where the series terminates at first order as $S$ is constructed by creation
operator $C^\dagger_I$ only \cite{coes,yx1}.
Evaluation of $\bar A_2|\Phi\rangle$ in general leaves only a constant and 
creation operators acting on $|\Phi\rangle$, namely,
\begin{equation}
\bar A_2|\Phi\rangle =\left[X_{2,0}(F;x_1,x_2)+\sum_I 
X_{2,I}(F;x_1,x_2)C^\dagger_I\right]|\Phi\rangle,
\end{equation}
$X_{2,0}$ and $X_{2,I}$ are a two-body function containing up to 
fourth-order terms in $F$. Therefore, the two-body density function 
in the Coester states is, using the definition of Eqs.~(11)
\begin{equation}
\rho^c_2(x_1,x_2)=X_{2,0}(F;x_1,x_2)+\sum_I X_{2,I}(F;x_1,x_2)\tilde g^c_I.
\end{equation}
In similar fashion, we derive
\begin{equation}
 \rho^c_n=X_{n,0}+\sum_I X_{n,I}\tilde g^c_I
\end{equation}
for the $n$-body density distribution function in the Coester states,
where $X_{n,0}$ and $X_{n,I}$ are the $n$-body functions
containing up to $(2n)$th-order terms in $F$. Therefore, the linked-cluster 
contributions of Eq.~(13) is written as
\begin{equation}
W^u=\frac12\int dx_1dx_2Z_2\left[X_{2,0}+\sum_IX_{2,I}\tilde g^c_I\right]+
\frac16\int dx_1dx_2dx_3Z_3\left[X_{3,0}+\sum_IX_{3,I}\tilde g^c_I\right]+\cdots,
\end{equation}
where the remainders are the 4-cluster and higher-order contributions,
and their calculations will depend on the details of the truncation schemes 
employed in the Coester states but the general property of finite-order 
in $F$ and linear in $\tilde g^c$ remains.

Before we consider the density-distribution functions of the Jastrow-Coester states 
of Eqs.~(1) and (2), we need  to define biased distribution functions as
\begin{equation}
\quad
 \tilde g_I\equiv\frac1{\langle\tilde\Psi|\Psi\rangle} 
 \langle\tilde\Psi^c|e^{S^0}C^\dagger_I|\Psi^c\rangle,\quad
 g_I\equiv\frac1{\langle\tilde\Psi|\Psi\rangle}
 \langle\tilde\Psi^c|C_Ie^{S^0}|\Psi^c\rangle.
\end{equation}
They are so called because they are not defined usually 
as $\langle C^\dagger_I\rangle$ and 
clearly $\tilde g_I\not=\langle C^\dagger_I\rangle$ due to 
the fact that $C^\dagger_I$ and $S^0/2$
do not commute in general. Similarly, $g_I\not=\langle C_I\rangle$
by our definition. These biased distribution functions can be calculated by
the functional derivative of the generating functional of Eq.~(8) as
\begin{equation}
\tilde g_I=\frac{\partial W}{\partial F_I}=\tilde g^c_I+\frac{\partial W^u}{\partial F_I},
\end{equation}
where we have used the definition of Eq.~(11). The functional derivative
in Eq.~(26) can be calculated by Eq.~(24),
\begin{equation}
 \frac{\partial W^u}{\partial F_I} =
 \frac12\int dx_1dx_2 Z_2\left\{\frac{\partial X_{2,0}}{\partial F_I}+\sum_{I'}\left[
 \frac{\partial X_{2,I'}}{\partial F_I}+X_{2,I'}(\tilde g^c_{I+I'}
   -\tilde g^c_I\tilde g^c_{I'})\right]\right\} + \cdots,
\end{equation}
where we have used the fact that
\begin{equation}
\frac{\partial\tilde g^c_{I'}}{\partial F_I} =
\frac{\partial^2 W^c}{\partial F_I\partial F_{I'}}=\tilde g^c_{I+I'}-\tilde g^c_I\tilde g^c_{I'},
\end{equation}
with $\tilde g^c_{I+I'}=\langle C^\dagger_IC^\dagger_{I'}\rangle^c=\langle C^\dagger_{I+I'}\rangle^c$.
Clearly $\tilde g^c_{I+I'}$ are also bare distribution functions.

Using the fact that $S^0/2$ commutes with density operator $\psi^\dagger(x)\psi(x)$,
the single-particle density function,
$\rho_1(x)=\langle \psi^\dagger(x)\psi(x)\rangle$ can then be calculated as
\begin{equation}
 \rho_1(x)=\frac1{\langle\tilde\Psi|\Psi\rangle}
   \langle\tilde\Psi^c|e^{S^0}\psi^\dagger(x)\psi(x)e^S|\Phi\rangle=
   \frac1{\langle\tilde\Psi|\Psi\rangle}
   \langle\tilde\Psi^c|e^{S^0}e^S\bar\psi^\dagger(x) \bar\psi(x)|\Phi\rangle
\end{equation}
where the evaluation of $\bar\psi^\dagger \bar\psi|\Phi\rangle =
(X_{1,0}+\sum_IX_{1,I}C^\dagger_I)|\Phi\rangle$ is similar to that in 
Eqs.~(19-23), hence we have, using the definition for the biased 
distribution function of Eq.~(25) and then using Eq.~(26),
\begin{equation}
\rho_1(x) = \rho^c_1(x)+ 
\sum_IX_{1,I}(F;x)\frac{\partial W^u}{\partial F_I},
\end{equation}
where $\rho^c_1(x)$ is the one-body density-distribution function 
of the Coester states as in general given by Eq.~(23) and the functional derivative
$\partial W^u/\partial F_I$ is given by Eq.~(27).

The two-body density distribution function of states of Eq.~(1) and (2) can be 
calculated in similar fashion as $\rho_1$ shown above. We take a more efficient
calculation by the functional derivative as
\begin{equation}
\rho_2(x_1,x_2)=\langle\psi^\dagger(x_1)\psi^\dagger(x_2)\psi(x_2)\psi(x_1)\rangle
 =2\frac{\partial W}{\partial u(x_1,x_2)}=2\frac{\partial W^u}{\partial u(x_1,x_2)}.
\end{equation}
The functional derivatives $\partial W^u/\partial u$ only involve 
the Yvon-Mayor functions $Z_n$ and it is easy
to derive from Eq.~(13)
\begin{equation}
\rho_2(x_1,x_2)=e^{u(x_1,x_2)}\rho^c_2(x_1,x_2)+
 \frac13\int dx'_1dx'_2dx'_3 \frac{\partial Z_3(x'_1,x'_2,x'_3)}
 {\partial u(x_1,x_2)}\rho^c_3(x'_1,x'_2,x'_3)+ \cdots,
\end{equation}
where $\rho^c_n$ are the $n$-body density-distribution function in the Coester states
as given in general by Eq.~(23). We can see immediately
from Eq.~(32) that the short-ranged correlation function $u(x_1,x_2)$ will
play an important role for applications to strongly-correlated systems, where the 
pure Coester states are known to be inefficient.

In summary, we have calculated the density-distribution functions as functionals 
of $(F,\tilde g^c,u)$ where $\tilde g^c$ are the bare density distribution functions 
discussed in our earlier VCCM papers \cite{yx1,yx2}; 
these calculations are all straightforward up to and including the third-order
cluster contributions as these functionals are all polynomials of $F$ and $\tilde g^c$,
and $u$ enters into these functionals through Yvon-Mayor functions.
The fourth- and higher-order cluster contributions will dependent on the details 
of applications with the truncation approximations employed in the Coester states. 
The first few terms of one-body and two-body functions are given by Eqs.~(30) and (32) 
respectively. We also want to point out that evaluation
of $\langle C^\dagger_I\rangle$ or $\langle C_I\rangle$ (or the 
off-diagonal density functions such as $\langle\psi^\dagger(x)\psi(x')\rangle$)
in general are highly nontrivial due to the fact that these operators
do not commute with the Jastrow factor $\exp(S^0/2)$. However, these
calculations are not needed in the Hamiltonian expectation value to be 
discussed in the next section; if necessary these calculations can be 
carried out in an approximation after we have obtained solutions to the 
variational equations \cite{rist}. We will discuss these calculations in future.

\section{Evaluation of Hamiltonian expectation value}

In evaluating a general Hamiltonian expectation value, we first notice 
that the kinetic part of Hamiltonian in general does not commute with the Jastrow 
operator $S^0/2$ in our states of Eqs.~(1) and (2). In real space, however, the 
kinetic operator contains only second-order derivatives in particle 
coordinates. We want to take this advantage by expressing our states in real space
for calculations. As shown in Appendix~A, the wavefunctions of the Jastrow-Coester 
states of Eqs.~(1) and (2) in real space are given by a product
\begin{equation}
\Psi = \Psi^c\Psi^u,\quad \tilde\Psi=\tilde\Psi^c \Psi^u
\end{equation}
where $\Psi^u=e^{U/2}=\exp[\sum_{i<j}u(x_i,x_j)/2]$ is the familiar Jastrow
wavefunction and $\Psi^c(x_1,\cdots,x_N)$ and $\tilde\Psi^c(x_1,\cdots,x_N)$ 
are real-space wavefunctions of the Coester ket- and bra-states respectively.
In general, we do not need to know the explicit functional form 
of $\Psi^c$ and $\tilde\Psi^c$ as our calculations involving them are
always carried out in a second quantization form as we 
show below. It is interesting nevertheless
to know that in a low-order SUB2 approximation, many-body function $\Psi^c$
is known explicitly as a partial-wave function, the so-called independent 
pair functions for boson gas \cite{fant2} or BCS superconductors \cite{carl}.
The Coester wavefunctions $\Psi^c$ and $\tilde\Psi^c$ obey proper symmetry, 
namely they are antisymmetric for fermions and symmetric for bosons under the 
exchange of any pair $x_i\rightleftharpoons x_j$. In the followings, we assume 
our states of Eqs.~(1) and (2) have a fixed particle number $N$ for convenience.
It is easy to extend to particle-number nonconserving states as discussed in
Appendix~A. Our final results are valid for both cases because they are 
expressed in terms of density distribution functions.

Evaluation of kinetic energy involving Jastrow function is helped
by Jackson-Feenberg transformation \cite{ripk}. We reproduce the
transformation as Eq.~(B7) in Appendix B for our wavefunctions of Eqs.~(33),
\begin{equation}
  \int dX\tilde\Psi\nabla^2_i\Psi = \int dX\left[\tilde\Psi^ce^U\nabla^2_i\Psi^c
  +\frac14\tilde\Psi^ce^U(\nabla^2_iU)\Psi^c-\frac14e^U\nabla^2_i(\tilde\Psi^c\Psi^c)\right].
\end{equation}
Another equivalent expression is also derived as Eq.~(B8),
\begin{equation}
 \int dX\tilde\Psi\nabla^2_i\Psi =
 \int dX\tilde\Psi^ce^U\left[\nabla^2_i+\frac14(\nabla^2_iU)
 +\frac12(\nabla_i U)\cdot\nabla_i\right]\Psi^c.
\end{equation}
Both transformations involve one- and two-body density distribution functions
only, and the Jastrow factor $e^U$ appears on the right of the
derivatives. The biased distribution functions defined in Eq.~(25) are then
applicable. The expectation value of a general Hamiltonian with an external 
field (and/or chemical potential) $A(x)$ is calculated as
\begin{equation}
E=\frac1{\langle\tilde\Psi|\Psi\rangle}
  \int dX\tilde\Psi\left[-\frac{\hbar^2}{2m}\sum_i\nabla^2_i 
  +\sum_i A(x_i)+\sum_{i<j}v(x_i,x_j)\right]\Psi.
\end{equation}
Using Eq.~(34), we derive
\begin{equation}
E_1=\int dx\left[-\frac{\hbar^2}{2m}\rho'_1(x)+\hat A_{eff}(x)\rho_1(x)\right]+
\frac12\int dx_1 dx_2v_{eff}(x_1,x_2)\rho_2,
\end{equation}
where $\rho_1$ and $\rho_2$ are the one- and two-body density distribution functions;
$\hat A_{eff}(x)$ is the effective external field operator
\begin{equation}
\hat A_{eff}=A(x)-\frac{\hbar^2}{8m}({\bf\nabla^c})^2
\end{equation}
with operator $\bf\nabla^c$ defined as applying to the Coester states only; 
$v_{eff}$ is the effective potential defined as
\begin{equation}
v_{eff}(x_1,x_2)= v(x_1,x_2)-\frac{\hbar^2}{8m}(\nabla_1^2+\nabla_2^2)u(x_1,x_2);
\end{equation}
and finally $\rho_1'$ is one-body density function derived from the first term of 
Eq.~(34) and written in the second quantization form as
\begin{equation}
\rho'_1(x)=\frac1{\langle\tilde\Psi|\Psi\rangle}\langle\tilde\Psi^c|e^{S^0}T_1|\Psi^c\rangle,
\quad T_1(x)= \psi^\dagger(x)\nabla^2\psi(x).
\end{equation}
Using Eq.~(35), we derive second equivalent energy functional as
\begin{equation}
E_2=\int dx\left[-\frac{\hbar^2}{2m}\rho'_1(x)+A(x)\rho_1(x)\right]+
\frac12\int dx_1 dx_2\left[v_{eff}(x_1,x_2)\rho_2-\frac{\hbar^2}{4m}\rho'_2(u)\right],
\end{equation}
where $\rho'_2(u)$ is two-body density functions derived from
the third integrals of Eq.~(35),
\begin{equation}
 \rho'_2(u)= \frac1{\langle\tilde\Psi|\Psi\rangle} 
 \langle\tilde\Psi^c|e^{S^0}T_2(u)|\Psi^c\rangle,
\end{equation}
with the two-body operator $T_2(u)$ given by
\begin{equation}
T_2(u) = \psi^\dagger(x_1)\psi^\dagger(x_2)
\left[\nabla_1 u(x_1,x_2)\cdot \nabla_1+\nabla_2u(x_1,x_2)\cdot\nabla_2\right]
  \psi(x_2)\psi(x_1).
\end{equation}

The difference between the two energy functionals is that in
$E_1$ of Eq.~(37) we need to take care of the operator $\bf\nabla^c$ which
applies only to the Coester states and in $E_2$ of Eq.~(41) we need to 
calculate the two-body density function $\rho'_2(u)$. We hope to get 
experience in real applications as which form is more practical. 
The density distribution functions $\rho_1$ 
and $\rho_2$ were calculated earlier by Eqs.~(30) and (32). It is easy to show 
that $\rho'_1$ and $\rho'_2(u)$ can be calculated in  similar fashion. We hence write
\begin{equation}
\rho'_1 =\frac1{\langle\tilde\Psi|\Psi\rangle}\langle\Psi^c|e^{S^0}T_1|\Psi^c\rangle
    =\frac1{\langle\tilde\Psi|\Psi\rangle}\langle\Psi^c|e^{S^0}e^S\bar T_1|\Phi\rangle
\end{equation}
where $\bar T_1(x)=e^{-S}T_1e^S=\bar \psi^\dagger(x)\nabla^2\bar \psi(x)$. After
evaluating $\bar T_1(x)|\Phi\rangle=[X'_{1,0}(F;x)+\sum_IX'_{1,I}(F;x)C^\dagger_I]|\Phi\rangle$
as before, we have
\begin{equation} 
 \rho'_1(x)=\rho'^c_1(x)+\sum_IX'_{1,I}(F;x)\frac{\partial W^u}{\partial F_I}
\end{equation}
where
\begin{equation}
\rho'^c_1(x)=\langle T_1\rangle^c =X'_{1,0}(F;x)+\sum_IX'_{1,I}(F;x)\tilde g^c_I
\end{equation}
is the similar density function of the Coester states,
and the derivative $\partial W^u/\partial F_I$ in Eq.~(45) is given by Eq.~(27).
In similar fashion, we derive
\begin{equation}
 \rho'_2(u;x_1,x_2)=\rho'^c_2+\sum_IX'_{2,I}\frac{\partial W^u}{\partial F_I}
 \end{equation}
where 
\begin{equation}
\rho'^c_2(u;x_1,x_2)=\langle T_2\rangle^c =X'_{2,0}+\sum_IX'_{2,I}\tilde g^c_I,
\end{equation}
and $X'_{2,0}$ and $X'_{2,I}$ are obtained by evaluating $\bar T_2(u)|\Phi\rangle= (X'_{2,0}+
\sum_IX'_{2,I}C^\dagger_I)|\Phi\rangle$.

Eqs.~(45) and (47), together with density distributions of Eq.~(30) and (32),
are all we need for calculating the two equivalent 
energy functional of Eq.~(37) and (41). These are our main results in this
paper. Denoting the three terms from the first term 
of $\rho'_1$, $\rho_1$ and $\rho_2$ as $\epsilon$, and the 
higher-order remainders as $\Delta E_l$
with $l=1,2$ for the two energy functionals, we 
rewrite the energy functionals of Eqs.~(37) and (41) as
\begin{equation}
E_l(F,\tilde g^c,u)=\epsilon(F,\tilde g^c,u)+\Delta E_l(F,\tilde g^c,u),\quad l=1,2
\end{equation}
where $\epsilon(F,\tilde g^c,u)$ is given by, 
\begin{equation}
\epsilon= K_1^c+\int dx A(x)\rho^c_1(x)+
\frac12\int dx_1dx_2 v_{eff}(x_1,x_2)e^{u(x_1,x_2)}\rho^c_{12}(x_1,x_2),
\end{equation}
with $K^c_1=-(\hbar^2/2m)\langle T_1\rangle^c$ for the kinetic energy of the VCCM
states. We notice that both energy functions give the same first order approximation
$\epsilon$ as the term containing $(\nabla^c)^2\rho^c_1(x)$ in $E_1$ vanishes 
after integration by parts. Furthermore, this first-order energy 
functional $\epsilon$ is nothing but the energy functional of the VCCM 
after replacing the bare potential $v\rightarrow V=v_{eff}e^u$, namely
\begin{equation}
\epsilon(F,\tilde g^c,u)=\left. E^c(F,\tilde g^c)\right|_{v\rightarrow V},\quad
 V(v,u)=\left[v(x_1,x_2)-\frac{\hbar^2}{8m}(\nabla_1^2
 +\nabla_2^2)u(x_1,x_2)\right]e^{u(x_1,x_2)}.
\end{equation}
This is convenient indeed as no new calculations are needed after the VCCM 
calculations have been done. We want to emphasize that, in addition to 
this simple, intuitively appealing approximation of Eq.~(51),
our main purpose here is to provide the detailed formulas for
calculating the higher-order terms in $\Delta E_l$ of Eq.~(49) in the forms
of four equations for $\rho_1$, $\rho'_1$, $\rho_2$
and $\rho'_2$ respectively.

\section{Summary and discussion}

In summary, we have calculated the Hamiltonian expectation
value of a general quantum many-body system as energy functionals of the 
Coester-state coefficients $\{F_I\}$, bare distribution 
functions $\{\tilde g^c_I\}$ of the Coester states and the Jastrow
correlation function $u(x_1,x_2)$. Two equivalent
energy functionals are derive by Eqs.~(37) and (41). A single simple, 
first-order energy from both expressions is derived as the usual VCCM 
energy functional but with a new potential $V(v,u)=v_{eff}e^u$. The formulas 
for practical calculations of the higher-order terms are provided in details. 
It is easy to see that, due to the short-ranged Jastrow factor, 
our energy functionals do not suffer the divergence as in 
most coupled-cluster calculations when potential $v(x_1,x_2)$ 
approaching hardcore potentials. We also want to emphasize that
an assumption has been made in applying the generalized linked-cluster
expansion of Eq.~(13) for the Coester states. We have proved this is valid
for a SUB2 truncation in the Coester states of spin lattice application
(and similar approximations for Bose gas and BCS state), we need to examine
this validity for the Coester states in a specific truncation
approximation for a real application. We believe this will not
pose major difficulty as we have the simple relations between the full 
distribution functions and the bare distribution functions of Eqs.~(23).
We also like to point out that the generalized linked-cluster expansion
of Eq.~(13) can not be applied to the traditional CCM due
to the linear construction of its bra state.

A similar wavefunctions to Eqs.~(33) were employed by Owen for study
of spin-dependent correlations in nuclear matter \cite{owen}. In particular,
$\Psi^c$ was approximated by the product of an independent pair function
(spin-dependent) and the Slater determinant; and in the cluster expansion, 
the Jastrow function and the independent pair function are treated together.
This differs from our approach. Our calculations involving Coester states are 
always in the second quantization form and are applicable to higher-order truncation
approximations. As mentioned in Introduction, KKZ \cite{krot2} employed the 
Coester ket state of Eq.~(1), and their energy and the equations for 
correlation coefficients were obtained by the Hamiltonian eigenequation. 
The validity of this procedure may be questionable as the Coester ket state
was approximated in a truncation approximation. In the traditional CCM approach, 
the use of Hamiltonian eigenequation is completely equivalent in any truncation 
approximation to a bi-orthogonal variational equations where the specific bra 
state of Eq.~(7) is  employed \cite{arpo,arpo2}, due to the fact
that the normalization integral is always unity and that the bra state is linear in
the coefficients. This is certainly not the case after inclusion of the Jastrow
function. In particular, the normalization integral in KKZ's 
equations must contain an additional term involving the bra-state coefficients
of Eq.~(7). It is not clear how this change will complicate their analysis.
However, we agree with KKZ and believe that the Jastrow theory and coupled-cluster theory
in principle complement each other and combination of these two theories may overcome each
other's difficulties. The approach presented here is our such attempt and
we have taken the advantage of the recent progress in the coupled-cluster
theory.

It is interesting to compare our energy functional
$E(F,\tilde g^c,u)$ with the counterpart in the traditional Jastrow theory, 
$E(a,u)$, where $a$ is the one-body function. Clearly, the missing 
state-dependent correlations in $E(a,u)$ are now included in $E(F,\tilde g^c,u)$
in terms of $F$ and $\tilde g^c$. However, a typical calculation of the Jastrow 
theory practiced today mostly includes resummation of all cluster terms
of the linked cluster expansion by the hyper-netted chain (hnc) approximation
for bosons or Fermi-hnc approximation for fermions \cite{fant,krot} and one 
optimization route for the boson system is provided by the pair-phonon analysis 
(PPA) of Campbell and Feenberg \cite{camp}. This is possible as the reference states 
are single-particle states. In our cluster expansion calculations, the reference
state is the Coester states which already contain rich correlations
including in particular correct long-ranged correlations such as in the SUB2
approximation. The introduction of the Jastrow function is
to provide correct description of strong, short-ranged
correlations and we expect such scheme may provide reasonable results even 
if we include a first few cluster terms. It will also be interesting
to investigate the relations between our approach and the PPA of
Campbell and Feenberg.

\acknowledgments
This work is part of our long-term research project. It is
not possible without the supports and encouragements from colleagues and
friends. In particular, this research was initially inspired by discussions 
with J. Arponen (deceased in 2006) and R.F. Bishop and by their works on an
extension of the CCM method \cite{arpo,arpo2}; we clearly share a goal to 
extend the traditional, well-established many-body techniques to a more 
general formalism capable of more accurate, consistent, complete description 
and applicable to a wider range of quantum many-body systems.

\begin{appendix}

\section{$\Delta$-representation and Generalized linked-cluster expansion}

We first consider the conversion relations between real-space wavefunctions
and the corresponding states in momentum-space. Let $|x_1,\cdots,x_N\rangle$ be 
the single-particle basis of a $N$-particle system, namely
\begin{equation}
|x_1,\cdots,x_N\rangle=\frac1{\sqrt{N!}}\psi^\dagger(x_N)\cdots\psi^\dagger(x_1)|0\rangle
\end{equation}
where $\psi^\dagger(x)$ are the field operators and $|0\rangle$ is the corresponding
vacuum state. The unity operator is then given by
\begin{equation}
1=\sum_N\int dX|x_1,\cdots,x_N\rangle\langle x_1,\cdots,x_N|,
\end{equation}
where $\int dX=\int dx_1\cdots dx_N$. The inner product between two arbitrary states
can be written as
\begin{equation}
\langle\Psi_1|\Psi_2\rangle =
 \sum_N\int dX\langle\Psi_1|x_1,\cdots,x_N\rangle\langle x_1,\cdots,x_N|\Psi_2\rangle
 =\sum_N\int dX \Psi^*_1\Psi_2,
\end{equation}
where $\Psi_1=\Psi_1(x_1,\cdots,x_N)=\langle x_1,\cdots,x_N|\Psi_1\rangle$
and $\Psi_2=\Psi_2(x_1,\cdots,x_N)=\langle x_1,\cdots,x_N|\Psi_2\rangle$ are the
corresponding wavefunctions in real space with proper boson or fermion
symmetry. If these two states have a fixed particle number $N$, 
the above inner product is then simply given by the familiar quantum mechanics formula
\begin{equation}
\langle\Psi_1|\Psi_2\rangle =\int dX \Psi^*_1 \Psi_2,
\end{equation}
because $\langle x_1,\cdots, x_{N'}|\Psi_2\rangle =\delta_{N',N}\Psi_2(x_1,\cdots,x_N)$.
In the followings, we assume our states have a fixed particle number $N$ for convenience.
It is easy to extend to particle-number nonconserving states by using Eq.~(A3). Our
final results in this paper are valid for both cases because they are expressed 
in terms of density distribution functions.

We denote $\Psi^c(x_1,\cdots,x_N)=\langle x_1,\cdots,x_N|\Psi^c\rangle$
as the Coester state function in real space, where Coester state $|\Psi^c\rangle$ is
given by Eq.~(4). The application of the Jastrow operator $S^0$ of Eq.~(3) 
on the single-particle basis can be written as
\begin{equation}
\frac{S^0}{2}|x_1,\cdots,x_N\rangle = \frac U2|x_1,\cdots,x_N\rangle,
\quad U = \sum_{i<j}u(x_i,x_j)
\end{equation}
or
\begin{equation}
e^{S^0/2}|x_1,\cdots,x_N\rangle = e^{U/2}|x_1,\cdots,x_N\rangle.
\end{equation}
We therefore have, using the fact that $u(x_i,x_j)$ is a real function,
\begin{equation}
\Psi(x_1,\cdots,x_N)=\langle x_1,\cdots,x_N|e^{S^0/2}|\Psi^c\rangle
  =e^{U/2}\langle x_1,\cdots,x_N|\Psi^c\rangle=\Psi^u \Psi^c,
\end{equation}
where $\Psi^u=e^{U/2}=\exp\left(\sum_{i<j}u(x_i,x_j)/2\right)$ is
the familiar Jastrow wavefunction. Similarly, the bra-state wavefunction
is written as
\begin{equation}
\tilde \Psi(x_1,\cdots,x_N)=\tilde\Psi^c\Psi^u=\tilde\Psi^ce^{U/2},
\end{equation}
where $\tilde\Psi^c$ are the corresponding bra-state wavefunctions of
Eqs.~(6) or (7) in real space. The Coester wavefunctions $\Psi^c$ and $\tilde\Psi^c$
obey proper symmetry, namely they are antisymmetric for fermions and symmetric 
for bosons under the exchange of any pair $x_i\rightleftharpoons x_j$.
We do not need to know their explicit functional forms as our later calculations
are always carried out in second quantization form of momentum space.

We follow the similar analysis for the evaluation of Eq.~(13) as in
the traditional Jastrow theory \cite{ripk,ripk2}. After the usual cluster expansion of the 
Jastrow wavefunction in terms of Yvon-Mayor functions $Z_n(Y)$ with
the bound function $Y_{12}=e^{u(x_1,x_2)}-1$,
\begin{equation}
e^U=1+\sum_{i<j}Z_2(x_i,x_j)+\sum_{i<j<k}Z_3(x_i,x_j,x_k)+\cdots,
\end{equation}
the expectation in Eq.~(13) is written as, 
\begin{equation}
\langle e^{S^0}\rangle^c =1+\frac1{I_c}\sum_{n=2}\frac{N!}{(N-n)!n!}\int dx_1\cdots dx_N
 \tilde \Psi^cZ_n(x_1,\cdots,x_n) \Psi^c=1+
 \sum_{n=2}\frac1{n!}\int dx_1\cdots dx_n Z_n\rho^c_n
\end{equation}
where $I_c=\langle\Psi^c|\Psi^c\rangle$ is the normalization integral,
the first few $Z_n$ are given by
\begin{equation}
Z_2=Y_{12},\quad Z_3=Y_{12}Y_{23}+Y_{23}Y_{31}+Y_{31}Y_{12}+Y_{12}Y_{23}Y_{31},
\end{equation}
etc., and $\rho^c_n$ is the $n$-body density distribution functions
\begin{equation}
\rho^c_n=\langle\psi^\dagger(x_1)\cdots\psi^\dagger(x_n)\psi(x_n)\cdots\psi(x_1)\rangle^c.
\end{equation}
Evaluation of $\rho^c_n$ of the Coester states in terms of bare distribution function
$\tilde g^c$ were discussed in details by Eqs.~(18-23) of Sec.~II.
Here we need to consider their cluster properties in order to apply
the linked-cluster theorem. For this purpose, we introduce the so-called
$\Delta$-representation discussed as follows. We consider evaluation of
$n$-body distribution functions in the state $|\Psi^d\rangle$ with the following
cluster decomposition property: any higher-order $\rho^d_n$ can be written as products
of lower order $\rho^d_m$ with $m=1,2,\cdots, n-1$, plus a nondecomposable
core $\Delta_n$. Let us demonstrate this property in details. We start with the 
one-body density distribution matrix, also the one-body core by definition,
\begin{equation}
\rho^d_1(x;x')=\langle\psi^\dagger(x)\psi(x')\rangle^d=\Delta_1(x;x'),
\end{equation}
where we have used the usual notation $\langle\cdots\rangle^d$ for the expectation
value in state $|\Psi^d\rangle$. The two-body distribution (diagonal) function, $\rho^d_2=\langle\psi^\dagger(x_1)\psi^\dagger(x_2)\psi(x_2)\psi(x_1)\rangle^d$
is then given by using this decomposition property,
\begin{equation}
\rho^d_2 =\Delta_1(x_1;x_1)\Delta_1(x_2;x_2) -\Delta_1(x_1;x_2)\Delta_1(x_2;x_1)
  +\Delta_2(x_1,x_2;x_1,x_2),
\end{equation}
where the first two terms are as given before for the fermions (for bosons, all terms
have positive sign) and the $\Delta_2$ is the 2-body
nondecomposable core. We can also write Eq.~(A14) in a symbolic notation as
\begin{equation}
\rho^d_2=\Delta_1*\Delta_1+\Delta_2.
\end{equation}
In similar way, the three-body density distribution function is then given by
\begin{equation}
\rho^d_3=\Delta_1*\Delta_1*\Delta_1+\Delta_1*\Delta_2+\Delta_3,
\end{equation}
where $\Delta_3$ is the nondecomposable core. All terms from the 
product $\Delta_1*\Delta_1*\Delta_1$ of Eq.~(A16) were given in details as 
diagrams in the references quoted earlier.
Similar to these diagrams, the new contributions in
products $\Delta_1*\Delta_2$ also include both unlinked terms such 
as $\Delta_1(x_1;x_1)\Delta_1(x_2,x_3;x_2,x_3)$, and linked terms such 
as $\Delta_1(x_1;x_3)\Delta_2(x_2,x_3;x_2,x_1)$. The rules for sign and for the
symbolic product $(*)$ can also all be defined in similar fashion
as before and we will discuss them in details somewhere else. Such decomposition
can be carried to higher-order.

Using the cluster decomposition property as discussed above, we have the key 
ingredient for the linked-cluster expansion, namely, all contributions in the 
product $Z_n\rho^d_n$ can be represented by clusters of diagrams. If we denote linked
diagrams as $\Gamma_A, \Gamma_B$, etc., a contribution in similar cluster expansion as
Eq.~(A10) for state $|\Psi^d\rangle$ can then be written as 
\begin{equation}
\Gamma=(\Gamma_A)^{\nu_A}(\Gamma_B)^{\nu_B}\cdots,\quad A\not= B,
\end{equation}
with coefficients 
\begin{equation}
N(A,B,\cdots) = \frac{n!}{(n_A!)^{\nu_A}(n_B!)^{\nu_B}\cdots}
\end{equation}
giving by the number of distributing the linked part along the $n$ points of
the diagram. We therefore have
\begin{equation}
\sum_{\Gamma_n}\frac1{n!}\Gamma_n=\sum_{A,B,\cdots}\frac1{n!}
N(A,B,\cdots)(\Gamma_A)^{\nu_A}(\Gamma_B)^{\nu_B}\cdots =
\exp(\Gamma_A/n_A!+\Gamma_B/n_B!+\cdots).
\end{equation}
Hence we have the following generalized linked-cluster expansion for the 
state $|\Psi^d\rangle$,
\begin{equation}
\ln \langle e^{S^0}\rangle^d=\sum_{n=2}\frac1{n!}\int dx_1\cdots dx_n [Z_n\rho^d_n]_L,
\end{equation}
where the notation $[Z_n\rho^d_n]_L$ denotes the contributions 
to $Z_n\rho^d_n$ by linked diagrams only. Since all terms 
in $Z_2$ and $Z_3$ are linked, the first two terms in Eq.~(A20) are 
independent of the diagram structures of $\rho^d_2$ and $\rho^d_3$ and we write
\begin{equation}
\ln \langle e^{S^0}\rangle^d=\frac12\int dx_1dx_2Z_2\rho^d_2+
\frac16\int dx_1dx_2dx_3 Z_3\rho^d_3+
\sum_{n=4}\frac1{n!}\int dx_1\cdots dx_n [Z_n\rho^d_n]_L.
\end{equation}
From $n=4$ and onward, we need to know the diagram details of $\rho^d_n$ for
calculating their contributions. 

In order to apply the generalized linked-cluster expansion of Eq.~(A21) to
our Coester states, we need to prove that the Coester states satisfy the 
cluster decomposition property as discussed above. In our earlier 
VCCM calculation for spin lattices, we have shown indeed the Coester
states satisfy such property in a SUB2 approximation employed, where
arbitrary order distribution functions can be calculated by the simple
functional derivative $\tilde g_{i'j'}/\partial F_{ij}=\tilde g_{ij'}\tilde g_{i'j}$ 
and these bare distribution functions correspond to the density distribution 
matrices (similar analysis also applied to the SUB2 state for the Bose gas 
and the BCS superconductors) \cite{yx2,yx5}. We also notice that a similar 
so-called SUB$m$ truncation approximation in the $\Delta$-representation can also
be defined as the approximation retaining up to $m$ core distribution tensors only.
It is intuitive to relate the real-space cluster parametrization by core distribution
tensor $\{\Delta_n\}$ in the $\Delta$-representation and momentum-space
parametrization by $\{F_I,\tilde g_I\}$ in the Coester states; 
the Coester representation provides a practical way to calculate these core 
tensors. We will not intend to provide a general proof that the Coester
states in any truncation approximation will satisfy the cluster decomposition
property. We will adopt a practical strategy and apply the the linked-cluster
expansion formula of Eq.~(A21) to the Coester states in real applications
and examine the cluster property in the particular truncation
approximation employed. We believe this will not cause a major difficulty 
as we use the relation between full distribution
functions $\rho_n^c$ and the bare distribution functions $\tilde g^c_I$
as given by Eqs.~(23).

We also like to point out that the traditional CCM states certainly
fail the cluster decomposition property due to the linear construction of
the bra state of Eq.~(7). This can be easily seen as any expectation in the
CCM is always linear in the bra state coefficients, contradictory to the
cluster decomposition property. Therefore, the generalized linked-cluster
expansion of Eq.~(A21) can not be applied to the CCM states.

\section{Jackson-Feenberg transformation}

We next consider the expectation value of the kinetic energy
operator using wavefunctions of Eqs.~(A7) and (A8). We follow the derivation
as given in Ref.~2 but keep using the notation $\tilde\Psi_C$
as it is different to the ket-state counterpart in the
traditional CCM. Applying the nested commutation formula in the following 
integral
\begin{equation}
I_i\equiv\int dX\tilde\Psi\nabla^2_i\Psi=
\int dX\tilde\Psi^c e^U\left(e^{-U/2}\nabla^2_ie^{U/2}\right)\Psi^c,
\end{equation}
we have
\begin{equation}
I_i=\int dX\tilde\Psi^ue^U\left[\nabla^2_i+\frac12(\nabla^2_i U)+(\nabla_i U)\cdot\nabla_i
 +\frac14(\nabla_i U)^2\right]\Psi^c.
\end{equation}
We can also apply the same nested commutation formula differently as,
\[
I_i=\int dX\tilde\Psi^c\left(e^{U/2}\nabla_ie^{-U/2}\right)e^U
  \left(e^{-U/2}\nabla_i e^{U/2}\right)\Psi^c,
\]
and derive another equivalent expression as
\begin{equation}
 I_i=\int dX\tilde\Psi^c\left[\nabla_i-\frac12(\nabla_i U)\right]e^U
  \left[\nabla_i+\frac12(\nabla_i U)\right]\Psi^c.
\end{equation}
Summing Eqs.~(B2) and (B3), the two $(\nabla_iU)^2$ terms (3-body terms) 
cancel each other out, we have
\begin{eqnarray*}
2I_i &=&\int dX\{\tilde\Psi^ce^U\nabla_i^2\Psi^c
  +\frac12\tilde\Psi^ce^U(\nabla_i^2 U)\Psi^c
  +\tilde\Psi^c\nabla_i e^U\nabla_i\Psi^c \\
  &+& \frac12\tilde\Psi^c\left[e^U(\nabla_i U)\nabla_i +
    \nabla_i e^U(\nabla_i U)\right]\Psi^c\},
\end{eqnarray*}
and integrating by parts for the third and last terms, we have
\begin{equation}
2I_i = \int dX\left[\tilde\Psi^ce^U\nabla^2_i\Psi^c
  +\frac12\tilde\Psi^ce^U(\nabla^2_iU)\Psi^c
  -(\nabla_i\tilde\Psi^c)e^U(\nabla_i\Psi^c)\right]+2B_i,
\end{equation}
where we have introduced $B_i$ as
\begin{equation}
B_i\equiv \frac14\int dX\left[\tilde\Psi^c e^U(\nabla_iU)\cdot(\nabla_i\Psi^c) 
    -e^U(\nabla_i\tilde\Psi^c)\cdot(\nabla_iU)\Psi^c\right].
\end{equation}
If $\tilde\Psi^c=\Psi^c$, these two 
terms cancel each other out and $B_i=0$, we therefore have
\begin{equation}
 I_i=\int dX\left[\frac12\tilde\Psi^ce^U\nabla^2_i\Psi^c
  +\frac14\tilde\Psi^ce^U(\nabla^2_iU)\Psi^c
  -\frac12(\nabla_i\tilde\Psi^c)e^U(\nabla_i\Psi^c)\right].
\end{equation}
Using the identity
\[
\nabla^2_i(\Psi^c)^2=2(\nabla_i\Psi^c)^2+2\Psi^c\nabla^2_i\Psi^c,
\]
the well-known Jackson-Feenberg formula is then derived as, with $\tilde\Psi^c=\Psi^c$,
\begin{equation}
 I_i= \int dX\left[\tilde\Psi^ce^U\nabla^2_i\Psi^c
  +\frac14\tilde\Psi^ce^U(\nabla^2_iU)\Psi^c-\frac14e^U\nabla^2_i(\tilde\Psi^c\Psi^c)\right].
\end{equation}
Eq.~(B6) can also be written as after integrating by part for the last
term,
\begin{equation}
 I_i=\int dX\tilde\Psi^ce^U\left[\nabla^2_i\Psi^c
  +\frac14(\nabla^2_iU)\Psi^c+\frac12(\nabla_iU)\cdot(\nabla_i\Psi^c)\right].
\end{equation}

We notice that in Eqs.~(B7) and (B8) we have assumed the Coester ket- and bra-state
wavefunctions equal, $\tilde\Psi^c=\Psi^c$. This is true for the VCCM states of
Eq.~(4-6) but not true in general for the CCM state of Eq.~(7) in
any finite truncation in $S$ and $\tilde S'$. Therefore the two terms in $B_i$
of Eq.~(B5) do not cancel in general in the the CCM basis. Our conclusion is 
that the kinetic expectation value in the CCM basis will also contain three-body 
density distribution functions.

\end{appendix}


\begin{references}
\bibitem{yx4} The variational equations in Refs. 20-21 were derived using energy 
functional $E(F,\tilde F)$. It is easy to show the equivalence of using 
energy functional $E(F,\tilde g^c)$.
\end{references}
\end{document}